\documentclass[11pt]{article}
\usepackage{latexsym}  
\usepackage{amssymb}
\usepackage{graphicx}
\usepackage{amsmath}

\begin{document}

\begin{center}
{\Large\bf Yukawaon Model with Anomaly Free Set\\

of Quarks and Leptons in a U(3) Family Symmetry}

\vspace{4mm}
{\bf Yoshio Koide$^a$ and Hiroyuki Nishiura$^b$}

${}^a$ {\it Department of Physics, Osaka University, 
Toyonaka, Osaka 560-0043, Japan} \\
{\it E-mail address: koide@kuno-g.phys.sci.osaka-u.ac.jp}

${}^b$ {\it Faculty of Information Science and Technology, 
Osaka Institute of Technology, 
Hirakata, Osaka 573-0196, Japan}\\
{\it E-mail address: nishiura@is.oit.ac.jp}

\date{\today}
\end{center}

\vspace{3mm}

\begin{abstract}
In the so-called ``yukawaon" model, the (effective) Yukawa coupling 
constants $Y_f^{eff}$ are given by vacuum expectation values (VEVs) of 
scalars $Y_f$ (yukawaons) with $3\times 3$ components.
So far, yukawaons $Y_f$ have been assigned to ${\bf 6}$ or ${\bf 6}^*$
of U(3) family symmetry, so that quarks and leptons were not anomaly
free in U(3).  In this paper, yukawaons are assigned to ${\bf 8}+{\bf 1}$
of U(3), so that  quarks and leptons are anomaly free. 
Since VEV relations among yukawaons are also  considerably changed, 
parameter fitting of the model is renewed.  After fixing our free 
parameters by observed mass ratios, we have only two and one 
remaining free parameters for quark and lepton mixings, respectively.  
We obtain successful predictions for the quark and lepton mixing parameters 
including magnitudes of $CP$ 
violation.  The effective Majorana neutrino mass is also predicted.
\end{abstract}

PCAC numbers:  
  11.30.Hv, 
  12.15.Ff, 
  14.60.Pq,  
  12.60.-i, 
\vspace{3mm}

\noindent{\large\bf 1 \ Introduction}

The central concern in the flavor physics is to understand 
masses and mixings of quarks and leptons. 
In this paper, we try to give a unified description of 
their mass spectra\footnote{
Note that in this paper, we investigate the origin of ``mass spectra", 
not the origin of ``masses"D
The origin of the ``masses" is elusive subject at any times.
The concept of ``mass" has been changed with the times.
It has, little by little, become clear as progress of the physics.
On the other hand, the origin of the ``mass spectra" has been a realistic 
subject at any times, and the investigation has played a
historical role in physics.
For the time being, we accept that the masses of quarks and leptons are 
generated by the Higgs mechanism.  
}
 and mixings based on the so-called ``yukawaon model"
\cite{yukawaon_PLB09,N-K_PRD11,K-N_EPJC12,K-N_PLB12,K-N_EPJC13,K-N_JHEP13} 
approach. 

First, we would like to give a brief review of the yukawaon model.

\vspace{3mm}

\noindent{\bf 1.1 \ What is a yukawaon?}

In the standard model (SM), the origin of mass spectra and mixing
is due to Yukawa coupling constants $Y_f$ ($f= u, d, \nu, e$), 
which are considered to be fundamental constants given
in physics and are incalculable. 
If we want to understand the families (generations) by a family symmetry,
we are obliged to regard the Yukawa coupling constants
as explicit symmetry breaking parameters, since we cannot construct such 
a model that is invariant under a non-Abelian 
family symmetry.

Against the view mentioned above, we think that 
the mass spectra and mixings are not fundamental quantities, 
but those that should be calculable dynamically.
Thus we regard the observed constants $Y^{eff}_f$ at a low energy scale as 
vacuum expectation values (VEVs) of scalars (yukawaons) $Y_f$:
$$
(Y_f^{eff})_i^{\ j} = \frac{y_f}{\Lambda} \langle (Y_f)_i^{\ j}\rangle ,
\eqno(1.1)
$$
where $\Lambda$ is a scale of the effective theory.

The conception of ``yukawaons" are summarized as follows:
(i) Yukawaons are a kind of flavons \cite{flavon}.
(ii) Those are singlets under the conventional gauge symmetries.
(iii) Since yukawaons are fields, we can consider a non-Abelian
family symmetry $G$ by assigning a suitable quantum number to $Y_f$.
(iv) The VEV forms are described by $3\times 3$ matrices. 
(v) Each yukawaon is distinguished from others by $R$ charges. 
(vi) VEV matrix relations are calculated from SUSY vacuum 
conditions.  
The relations are given by form of a product of VEV matrices
(not  form of sum of those) as we show later.   
(vii) The VEV matrix $\langle Y_f \rangle$ also evolves after the 
symmetry breaking in the same way that a conventional Yukawa coupling constant in the SM does.

In the yukawaon model, $R$ charge assignments are essential for 
obtaining successful phenomenological results. 
Although we assign $R$ charges from the phenomenological 
point of view, the assignments cannot be taken freely. 
We must take the assignments so that they may forbid  
appearance of unwelcome terms. 
(The details are discussed in Sec.2.4 later.)
     
VEV matrix relations among yukawaons are obtained as follows:
First, we write down a superpotential which is invariant under 
the family gauge symmetry $G$ (we will consider $G$=U(3) 
in the present model) with considering $R$ charge conservation. 
Next, we apply a SUSY vacuum condition to the superpotential 
to get a VEV matrix relation.
For example, from a SUSY vacuum condition
$\partial W_R/\partial \Theta_R =0$ for 
a superpotential $W_R$ which is given by 
$$
W_R = \mu_R  (Y_R)_{i j} \Theta^{j i}_R 
+\lambda_R \left( (Y_e)_i^{\ k} (\Phi_u)_{k j}
+ (\Phi_u)_{i k} (Y_e^T)^k_{\ j} \right) \Theta^{j i}_R ,
\eqno(1.2)
$$
we obtain a VEV relation
$$
\langle Y_R \rangle = -\frac{\lambda_R}{\mu_R} \left( \langle Y_e \rangle 
\langle \Phi_u \rangle + \langle\Phi_u\rangle \langle Y_e\rangle  
\right) .
\eqno(1.3)
$$ 
Here $\Phi_u$ is a subsidiary flavon whose VEV is related to 
a VEV of the up-quark yukawaon $Y_u$ as
$\langle Y_u \rangle = k_u \langle\Phi_u\rangle \langle \bar{\Phi}_u\rangle$ 
as we see later.
For the time being, we assume that the observed supersymmetry breaking 
is induced by a gauge mediation mechanism (not including family 
gauge symmetries), 
so that our VEV relations among yukawaons are still valid after
the SUSY was broken in the quark and lepton sectors.

\vspace{3mm}

\noindent{\bf 1.2 \ What is our aim?}

It is an attractive idea that observed hierarchical 
structures of masses and mixings of quarks and leptons
are caused by a common origin. 
We suppose that the observed hierarchical family structures 
are  caused by 
one common origin, so that they can successfully be understood
by accepting one of the hierarchical structures (e.g. 
charged lepton mass spectra) as input values of the model.
If it is true, we will be able to describe quark and lepton 
mass matrices without using any other family-number dependent 
input parameters, except for the observed values of 
the charged lepton masses $(m_e, m_\mu, m_\tau)$ as input parameters 
with hierarchical values.  
Here, the terminology ``family-number independent parameters" is used, for 
example, for the coefficients of a unit matrix ${\bf 1}$, a democratic matrix $X_3$, 
and so on, where
$$
{\bf 1} = \left( 
\begin{array}{ccc}
1 & 0 & 0 \\
0 & 1 & 0 \\
0 & 0 & 1 
\end{array} \right) , \ \ \ \ \ 
X_3 = \frac{1}{3} \left( 
\begin{array}{ccc}
1 & 1 & 1 \\
1 & 1 & 1 \\
1 & 1 & 1 
\end{array} \right) .
\eqno(1.4)
$$
(For an explicit example of the previous yukawaon model, 
see Eqs.(1.5) and (1.6) for example.) 
Regrettably, at present, we are obliged to accept to use a few other family-number 
dependent parameters.  
For example, we still use a phase matrix $P= {\rm diag}
(e^{i\phi_1},  e^{i\phi_2}, 1)$ 
in the Cabibbo-Kobayashi-Maskawa (CKM)  quark mixing matrix 
$V_{CKM}=U_u^\dagger P\, U_d$ \cite{CKM}. 
Therefore, our original intention in the yukawaon model is not yet 
completed at present. 

Even if we finally fail to describe quark and lepton mass matrices without 
any family-number dependent parameters, it only means that the observed 
hierarchical structures of quarks and lepton masses and mixings 
are caused by two origins.
In either case, it is important as the first step to investigate a common origin for the hierarchy.

\vspace{3mm}

\noindent{\bf 1.3 \ Past yukawaon models}

In the earlier stage of yukawaon models \cite{yukawaon_PLB09,N-K_PRD11},  
the VEV matrices of yukawaons 
have been described  by the following quark and lepton mass matrices:
$$
\begin{array}{l}
Y_e = k_e \Phi_e \Phi_e,  \\ 
Y_\nu = Y_D Y_R^{-1} Y_D^T , \\
Y_u = k_u \Phi_u \Phi_u, \\   
Y_d = k_d P_d \Phi_e ( {\bf 1} + a_d X_3) \Phi_e P_d , \\ 
\end{array}
\eqno(1.5)
$$
with subsidiary conditions
$$
\begin{array}{l}
\Phi_e= k'_e {\rm diag}(\sqrt{m_e}, \sqrt{m_\mu}, \sqrt{m_\tau})    \\
\Phi_u = k'_u \Phi_e ( {\bf 1} + a_u X_3) \Phi_e , \\
Y_D = Y_e , \\
Y_R = k_R (\Phi_u Y_e + Y_e \Phi_u ) + \cdots ,
\end{array}
\eqno(1.6)
$$
where $Y_e$,  $Y_u$, $Y_d$, $Y_\nu$, $Y_D$, $Y_R$, and $Y_\nu$ 
correspond to charged lepton, up-quark, down-quark, neutrinos, Dirac neutrino, 
Majorana right-handed neutrino mass matrices, respectively, 
and  $P_d$ is a phase matrix $P_d = {\rm diag} (e^{i \phi_1}, e^{i \phi_2}, 1)$.  
(Here, we have denoted a VEV matrix $\langle Y_f \rangle$ as $Y_f$ simply.)
The coefficients $a_u$ and $a_d$ are family-number independent parameters.
On the other hand, since we discuss only mass ratios and mixings, 
the parameters $k_e$, $k_u$ and so on are not essential in the model. 
(Hereafter, we omit such common coefficients.)

In the VEV matrix relations, a factor $({\bf 1}+a_f X_3)$ plays an essential
role. 
We assume an existence of the following flavor basis: (i) A fundamental flavon
VEV matrix $\langle \Phi_e\rangle$ (we denote $\langle \Phi_0\rangle$ later)
is diagonal; (ii) On this flavor basis, the VEV matrix $\langle X_3 \rangle$ takes 
a democratic form $X_3$ defined by (1.4).
(Such the flavor basis has been assumed in a ``democratic universal seesaw model" 
\cite{K-F_ZPC96}.) 

The model (1.5) with (1.6) could give tribimaximal mixing for 
the Pontecorvo-Maki-Nakagawa-Sakata (PMNS) lepton mixing matrix 
$U_{PMNS}$ \cite{PMNS},
but it gave poor fitting for $V_{CKM}$.
Besides, the model could not give the observed large mixing \cite{theta13}
$\sin^2 2\theta_{13} \sim 0.09$, whose value was reported  
after the proposal of the model (1.5).

In the second stage of the yukawaon model \cite{K-N_EPJC12}, 
stimulated by this new observation $\sin^2 2\theta_{13} \sim 0.09$, 
we proposed to change the previous structure 
$Y_e =k_e \Phi_e \Phi_e$ 
into a new structure
$$
Y_e = k_e \Phi_0 ( {\bf 1} + a_e X_3 ) \Phi_0. 
\eqno(1.7)
$$
Here, similar to $\Phi_d$ given in Eq.(1.5), a  VEV matrix $\Phi_0$ is given by 
a diagonal matrix $\Phi_0 = {\rm diag}(x_1, x_2, x_3)$ and 
the parameters $(x_1, x_2, x_3)$ are only hierarchical parameters 
in this model.
Then the charged lepton mass matrix $Y_e$ is not diagonal
any longer.
This alteration means a serious change for the yukawaon model, 
because one of the purposes in the yukawaon model was to 
understand a charged lepton mass relation \cite{Koidemass}
$(m_e + m_\mu + m_\tau)/(\sqrt{m_e} + \sqrt{m_\mu} +
\sqrt{m_\tau} )^2 =2/3$ by using the relation 
$Y_e =k_e \Phi_e \Phi_e$.  
The new form of $Y_e$, Eq.(1.7) cannot lead to the charged 
lepton mass relation any more. 
 
Also, the assumption $Y_D=Y_e$ was changed into
$$
Y_D = \Phi_0 ( {\bf 1} + a_D X_2 ) \Phi_0 \neq Y_e ,  
\eqno(1.8)
$$
where
$$
X_2 = \frac{1}{2} \left( 
\begin{array}{ccc}
1 & 1 & 0 \\
1 & 1 & 0 \\
0 & 0 & 0 
\end{array} \right).  
\eqno(1.9)
$$ 
Furthermore, recently, we have proposed \cite{K-N_JHEP13} a new neutrino 
mass matrix with a bilinear form
$Y_\nu = (Y_D Y_R^{-1} Y_D)^2$.
The model can give reasonable predictions of the 
quark and lepton mixings ($V_{CKM}$ and $U_{PMNS}$) 
together with their masses, but we still failed 
to give a large value of $\sin^2 2\theta_{13} \sim 0.09$. 

\vspace{3mm}

\noindent{\bf 1.4 \ What is new?}

The purpose of the present paper is not to improve parameter fitting, but to improve 
a basic part of the yukawaon model. 
Of course, we will give a reasonable parameter fitting
to the observables including a large value of $\sin^2 \theta_{13}$ 
by using a new model in which the number of free parameters is less than those in the 
previous works. 
So far, the yukawaons $Y_f$ in the previous works have been described as 
$Y_f^{ij}$, i.e. ${\bf 6}^*$ of a family symmetry U(3).
The reason is as follows: 
If we consider a field $C$ of ${\bf 8}+ {\bf 1}$ of U(3),
i.e. $C_i^{\ j}$, and we require a triple product $A C B$ 
by $A^{ik} C_k^{\ l} B_{lj}$, then, we are obliged 
to have unwelcome triple products $A^{ik}  B_{kl} (C^T)^l_{\ j}$
and $ (C^T)^i_{\ k} A^{kl} B_{lj}$. 
In the yukawaon model, the order of multiplication of matrices is essential.
Therefore, so far, we have not adopted a yukawaon model with 
$(Y_f)_i^{\ j}$.
However, in the model with $(Y_f)^{ij}$, quarks and leptons $f$ 
are assigned to $(f_L, f_R) \sim ({\bf 3}, {\bf 3}^*)$ 
of the U(3) family symmetry, so that the fundamental fermions 
are not anomaly free in the U(3) symmetry.  
In the present model, the yukawaons are given by $(Y_f)_i^{\ j}$, 
so that quarks and leptons, themselves, are anomaly free for 
the U(3) family symmetry.
Of course, there is no reason that quarks and leptons must 
compose an anomaly free set. 
Alternatively, in the previous model, we have assumed a supersymmetric theory 
(SUSY), so that the model could become anomaly free by taking 
whole flavons in the model into consideration, although 
quarks and leptons, themselves, were not anomaly free. 
However, we empirically know that quarks and leptons, which are fundamental
entities in the low energy limit (in the standard model limit), 
compose an anomaly free set of gauge symmetries concerned.
So it is natural that quarks and leptons compose an anomaly free set in 
family gauge symmetry U(3), too.

In this paper, the following points are renewed:

\noindent (i)  As we have stated above, we use new yukawaons $(Y_f)_i^{\ j}$ 
instead of $(Y_f)^{ij}$ which are used in the past models. 

\noindent (ii) A seesaw mass matrix for the neutrino mass matrix 
$M_\nu =(M_D M_Y^{-1} M_D)^2$ is explicitly given by an extended seesaw 
mechanism.  (See Eq.(2.12) later.)

\noindent (iii)  VEV of the yukawaons $Y_f$ in the quark sectors 
($f=u,d$) are given by a bilinear form $Y_f= \Phi_f \Phi_f$, 
while yukawaon VEVs in lepton sectors $Y_e$ and $Y_D$ are given 
by the form $Y_f = \Phi_f$:
$$
Y_e = \Phi_e, \ \ \ Y_D = \Phi_D, \ \ \ Y_u = \Phi_u \Phi_u, 
\ \ \ Y_d = \Phi_d \Phi_d .
\eqno(1.10)
$$

\noindent (iv) Correspondingly to the change from $(Y_f)^{ij}$ 
to $(Y_f)_i^{\ j}$, the flavon $\Phi^{ij}_0$ of U(3) is changed 
into two kind of $\Phi_0$, $(\Phi_0)_{i\alpha}$ and 
$(\bar{\Phi}_0)^{i\alpha}$, 
which are $({\bf 3}, {\bf 3})$ and $({\bf 3}^*, {\bf 3}^*)$
of U(3)$\times$U(3)$'$, respectively.
(More details are given in the next section.)

In the next section, we give a renewed yukawaon model with $({Y}_f)_i^{\ j}$.
We give VEV matrices relations among 
flavons (yukawaons). 
Those flavons are distinguished by the $R$ charges 
assignments of which are discussed in Sec.2.4.
Renewed parameter fitting is given in Sec.3.
Finally, Sec.4 is devoted to concluding remarks.

\vspace{3mm}

\noindent{\large\bf 2 \ Model}

\vspace{2mm}

\noindent{\bf 2.1 \ Overview of the model}

Hereafter, for convenience, we denote a flavon $A$ with ${\bf 6}^*$ 
as $\bar{A}$, and a flavon $A$ with ${\bf 8} +{\bf 1}$ as $\hat{A}$.\\

{\bf A. \ Would-be Yukawa interactions}

We assume that a would-be Yukawa interaction is given as follows:
$$
W_Y = \frac{y_D}{\Lambda} (\nu^c)^i (\hat{Y}_D^T)_i^{\ j} \ell_j  H_u 
+ \frac{y_e}{\Lambda} (e^c)^i (\hat{Y}_e)_i^{\ j}\ell_j  H_d 
+ \frac{y_u}{\Lambda}  (u^c)^i (\hat{Y}_u)_i^{\ j} q_j H_u 
+ \frac{y_d}{\Lambda}  (d^c)^i (\hat{Y}_d)_i^{\ j} q_{j} H_d  
$$
$$
+y_R (\nu^c)^i (Y_R)_{ij} (N^c)^j   
+ y'_D (N^c)^i (\hat{Y}_D)_i^{\ j} N_j  
+ y_N N_i (\bar{E}_N)^{ij} N_j ,
\eqno(2.1)
$$
where $\ell=(\nu_L, e_L)$ and $q=(u_L, d_L)$ are SU(2)$_L$ doublets,  
and $N$ and $N^c$ are new  SU(2)$_L$ singlet leptons. 
The last three terms in Eq.(2.1) are added in order to give the 
neutrino mass matrix with a form $M_\nu =(Y_D Y_R^{-1} Y_D)^2$
as we show later.  
  
In order to distinguish each yukawaon from others, we assume that 
yukawaons $Y_f$ have different $R$ charges\footnote{
If we assume a U(1) charge conservation (but not the $R$ charges) 
in order to distinguish 
those yukawaon, each term in the superpotential is assigned to have a charge $Q=0$.
Then, for arbitrary two of those terms (say terms $A$ and $B$),
the product $A\cdot B$ is also has $Q=0$, so that $A\cdot B$ is
allowed as an additional term in the superpotential.  
On the other hand, when we assume $R$ charge conservation, 
$A$ and $B$ have a charge $R=2$, the term $A\cdot B$ has 
$R=4$, so that the term  $A\cdot B$ is forbidden in the superpotential.
} 
from each other with considering $R$ 
charge conservation (a global U(1) symmetry in $N=1$ supersymmetry).
(Of course, the $R$ charge conservation is broken
at an energy scale $\Lambda$, at which the U(3) family symmetry 
is broken.)
Possible assignments of $R$ charges of the flavons are given 
in Sec.2.4. 

Let us comment on $R$ parity assignments.
Since we inherit $R$ parity assignments in the standard SUSY model,
$R$ parities of yukawaons $Y_f$ (and all flavons) are the same 
as those of Higgs particles 
(i.e. $P_R({\rm fermion})=-1$ and $P_R({\rm scalar})=+1$), 
while quarks and leptons are assigned to 
$P_R({\rm fermion})=+1$ and $P_R({\rm scalar})=-1$. \\

{\bf B. \ VEV relations among flavons}

Each yukawaon has the
basic structure $\Phi_0 ({\bf 1} + a_f X_3)\Phi_0$ as well as the previous 
models.
Explicitly speaking, a VEV $\langle \bar{Y}_e \rangle$ of the charged lepton 
yukawaon takes a basic structure 
$$
\langle\hat{Y}_e\rangle_i^{\ j} = 
\langle {\Phi}_0 \rangle_{i\alpha} 
\left( \langle (\bar{E}_0)\rangle^{\alpha \gamma} 
\langle {E}_0 \rangle_{\gamma \beta}  
+ a_e  \langle (\bar{X}_3) \rangle^{\alpha \gamma} 
\langle {X}_3 \rangle_{\gamma \beta}  \right) 
\langle \bar{\Phi}_0^T \rangle^{\beta j}  ,
\eqno(2.2)
$$
by SUSY vacuum conditions. 
VEVs of the up- and down-quark yukawaons, $ \langle\hat{Y}_u\rangle$ and 
$\langle \hat{Y}_d\rangle$ are given by 
$$
\langle \hat{Y}_u \rangle_i^{\ j} = \langle {\Phi}_u \rangle_{i k}
\langle \bar{\Phi}_u \rangle^{k j} , \ \ \ \ 
\langle \hat{Y}_d \rangle_i^{\ j} = \langle {\Phi}_d \rangle_{i k}
\langle \bar{\Phi}_d \rangle^{k j} ,
 \eqno(2.3)
$$
where VEVs of flavons ${\Phi}_u$, $\bar{\Phi}_u$, ${\Phi}_d$ and
$\bar{\Phi}_d$ are given by 
$$
\begin{array}{l}
\langle {\Phi}_u\rangle_{i j} = 
\langle P_u \rangle_{ik} \langle \bar{\Phi}_0 \rangle^{k\alpha} 
\left( \langle {E}_0 \rangle_{\alpha \beta} + a_u 
 \langle {X}_3 \rangle_{\alpha \beta} \right) 
\langle \bar{\Phi}_0^T\rangle^{\beta l}  \langle {P}_u \rangle_{lj} , \\
\langle \bar{\Phi}_u\rangle^{i j} = 
\langle \bar{P}_u \rangle^{ik} \langle {\Phi}_0 \rangle_{k\alpha} 
\left( \langle \bar{E}_0 \rangle^{\alpha \beta} + a_u 
 \langle \bar{X}_3 \rangle^{\alpha \beta} \right)  
\langle {\Phi}_0^T\rangle_{\beta l}  \langle \bar{P}_u \rangle^{lj} , 
\end{array}
\eqno(2.4)
$$
$$
\begin{array}{l}
\langle {\Phi}_d\rangle_{i j} = 
\langle E_d \rangle_{ik} \langle \bar{\Phi}_0 \rangle^{k\alpha} 
\left( \langle {E}_0 \rangle_{\alpha \beta} + a_d 
 \langle {X}_3 \rangle_{\alpha \beta} \right) 
\langle \bar{\Phi}_0^T\rangle^{\beta l}  \langle {E}_d \rangle_{lj} , \\
\langle \bar{\Phi}_d\rangle^{i j} = 
\langle \bar{E}_d \rangle^{ik} \langle {\Phi}_0 \rangle_{k\alpha} 
\left( \langle \bar{E}_0 \rangle^{\alpha \beta} + a_d 
 \langle \bar{X}_3 \rangle^{\alpha \beta} \right)  
\langle {\Phi}_0^T\rangle_{\beta l}  \langle \bar{E}_d \rangle^{lj} .
\end{array}
\eqno(2.5)
$$
Here, we have dropped common coefficients which do not affect relative 
relations among families.

We take 
$$
\langle \Phi_0 \rangle = \langle \bar{\Phi}_0 \rangle = 
{\rm diag} (x_1, x_2, x_3) ,
\eqno(2.6)
$$
from the $D$-term condition, where $x_i$ are real.
In general, for VEV matrices $\langle A\rangle$ and 
$\langle \bar{A} \rangle$, we can choose either one in two cases
$$
\langle \bar{A} \rangle = \langle {A} \rangle^{*} ,
\eqno(2.7)
$$
$$
\langle \bar{A} \rangle = \langle {A} \rangle .
\eqno(2.8)
$$
We have assumed the case (2.8) for the relation (2.3), 
while we have applied the case (2.7) to the VEV matrices  
$\langle P_u\rangle$ and $\langle \bar{P}_u \rangle$,
i.e. 
$$
\langle P_u \rangle = {\rm diag}( e^{i\phi_2}, e^{i\phi_2},1), \ \ \ 
\langle \bar{P}_u \rangle = {\rm diag}( e^{-i\phi_2}, e^{-i\phi_2},1).
\eqno(2.9)
$$

{\bf C. \ Neutrino sector}

Finally, let us explain 
the neutrino mass matrix which is given by $M_\nu =(Y_D Y_R^{-1} Y_D)^2$.  
From the Yukawa interactions (2.1), we can write a mass matrix for 
neutral leptons $(\nu, \nu^c, N^c, N)$ as follows:
$$
M_{4\times 4}= \left(
\begin{array}{cccc}
0 & \frac{y_D v_H}{\Lambda} \langle\hat{Y}_D^T\rangle & 0 & 0 \\
\frac{y_D v_H}{\Lambda} \langle\hat{Y}_D\rangle & 0 & 
y_R \langle Y_R\rangle & 0 \\
0 & y_R \langle Y_R \rangle & 0 & y'_D \langle\hat{Y}_D\rangle \\
0 & 0 & y'_D \langle\hat{Y}_D^T\rangle & y_N \langle E_N \rangle
\end{array} \right) .
\eqno(2.10)
$$
We apply a triplicated seesaw approximation to Eq.(2.10) as follows:
$$
M_{4\times 4} \ \Rightarrow \ M_{3\times 3} \simeq  \left(
\begin{array}{ccc}
0 & \frac{y_D v_H}{\Lambda} \langle\hat{Y}_D^T\rangle & 0  \\
\frac{y_D v_H}{\Lambda} \langle\hat{Y}_D\rangle & 0 & 
y_R \langle Y_R\rangle  \\
0 & y_R \langle Y_R \rangle & -y'_D \langle\hat{Y}_D^T\rangle 
( y_N \langle E_N \rangle)^{-1} y'_D \langle\hat{Y}_D\rangle 
\end{array} \right)
$$
$$
\Rightarrow   M_{2\times 2} \simeq   \left(
\begin{array}{cc}
0 & \frac{y_D v_H}{\Lambda} \langle\hat{Y}_D^T\rangle  \\
\frac{y_D v_H}{\Lambda} \langle\hat{Y}_D\rangle & 
y_R \langle Y_R\rangle [-y'_D \langle\hat{Y}_D^T\rangle 
( y_N \langle E_N \rangle)^{-1} y'_D \langle\hat{Y}_D\rangle ]^{-1}
y_R \langle Y_R\rangle
\end{array} \right) \ \Rightarrow \ M_{1\times 1} ,
\eqno(2.11)
$$
where
$$
M^{ij}_\nu \equiv  (M_{1\times 1})^{ij}  
\simeq  \langle\hat{Y}_D^T\rangle^i_{\ k} \langle Y_R^{-1}\rangle^{kl} 
\langle\hat{Y}_D\rangle_l^{\ m} 
\langle E_N^{-1}\rangle_{mm'} \langle\hat{Y}_D^T\rangle^{m'}_{\ l'} 
\langle Y_R^{-1}\rangle^{l'k'}
\langle\hat{Y}_D\rangle_{k'}^{\ j} .
\eqno(2.12)
$$
In the expression (2.12), we have dropped common coefficients which
do not affect relative ratios among families.   
Here we have assumed 
$$
y_N^2 |\langle E_N\rangle|^2 \gg y_R^2 |\langle Y_R \rangle|^2 \gg
y_D^2 |\langle \hat{Y}_D \rangle|^2 ,
\eqno(2.13)
$$
in order to obtain good seesaw approximation (2.11). 
We consider that there are hierarchical structures 
not only among the VEV values of $|\langle E_N\rangle|^2$, 
$|\langle Y_R \rangle|^2$ and $|\langle \hat{Y}_D \rangle|^2$, 
but also among the coupling constants $y_N^2$, $y_R^2$ and $y_D^2$. 
This type of the neutrino mass matrix is known as ``inverse seesaw" 
model \cite{inverse_seesaw}

Here, we assume that the Yukawaon VEV matrix $\langle \hat{Y}_D \rangle$ for  
Dirac neutrinos takes a
 slightly different form from that for $Y_e$:
$$
\langle\hat{Y}_D\rangle_i^{\ j} = 
\langle E_D \rangle_{i \alpha}
\langle \bar{\Phi}_0^T \rangle^{\alpha k} 
\left( \langle \hat{E}'_0 \rangle_k^{\ l} 
+ a_D  \langle \hat{X}_2 \rangle_k^{\ l} \right) 
\langle {\Phi}_0 \rangle_{l \beta} 
\langle \bar{E}_D \rangle^{\beta j}  .
\eqno(2.14)
$$
Note that against the universal form Eqs.(2.2), (2.4) and (2.5), 
the matrix $X_3$ has been replaced with $X_2$ in Eq.(2.14). 
The form $X_2$ has been brought in the present model from the 
phenomenological reason, and the form is ad hoc one.   
Although we speculated  a mechanism  \cite{K-N_EPJC13} for the form 
$\bar{X}_2$ by introducing additional family symmetry U(3)$'$, 
we do not refer to such a mechanism in the present paper.
The origin of the structure $\bar{X}_2$ is left to our future task. 

On the other hand, the Majorana neutrino mass matrix $\langle Y_R\rangle$ is
given by
$$
 \langle Y_R\rangle_{ij}  = \langle\hat{Y}_e\rangle_i^{\ k} 
\langle{\Phi}_u\rangle_{kj} + 
\langle{\Phi}_u\rangle_{ik}  \langle\hat{Y}_e^T \rangle^k_{\ j}  .
\eqno(2.15)
$$ 

Superpotential forms which lead to the VEV relations mentioned above and $R$ 
charge assignments are given in Sec.2.2 and 2.4. 
Specific forms $E$, $P$ and $X_3$ are discussed in Sec.2.3. 

\vspace{2mm}

\noindent{\bf 2.2 \ Superpotential and VEV relations}

The VEV matrix relations given in Sec.2.1 are obtained from 
SUSY vacuum conditions.
For example, the VEV relation (2.2) can be obtained by 
requiring a SUSY vacuum condition $\partial W/\partial \Theta_e=0$
for the following superpotential: 
$$
W_e= \left\{ \mu_e  (\hat{Y}_e)_i^{\ j} + 
\frac{\lambda_e}{\Lambda^3} (\bar{\Phi}_0)_{i\alpha}
\left(  (\bar{E}_0)^{\alpha \gamma} ({E}_0)_{\gamma \beta}  + a_e 
 (\bar{X}_3)^{\alpha \gamma} ({X}_3)_{\gamma \beta}   \right) 
 ({\Phi}_0^T)^{\beta j} \right\} 
  (\hat{\Theta}_e)_{j}^{\ i}.
\eqno(2.16)
$$
Since we assume that the $\Theta$ field always takes 
 $\langle \Theta \rangle =0$ and since SUSY vacuum conditions 
in other fields always contain the VEV matrix  
$\langle \Theta \rangle $, such conditions do not play 
any effective role in obtaining VEV relations.

Similarly, the VEV relations (2.3), (2.14) and (2.15) are obtained 
from the following superpotential:
$$
W_u= \left\{ \mu_u (\hat{Y}_u)_i^{\ j} +
{\lambda_u} 
({\Phi}_u)_{ik} (\bar{\Phi}_u )^{k j}
\right\} (\hat{\Theta}_u)_j^{\ i} , \\
\eqno(2.17)
$$
$$
W_d= \left\{ \mu_d (\hat{Y}_d)_i^{\ j} +
{\lambda_d} 
({\Phi}_d)_{ik} (\bar{\Phi}_d )^{k j}
\right\} (\hat{\Theta}_d)_j^{\ i}  , 
\eqno(2.18)
$$
$$
W_D=  \left\{ \mu_D (\hat{Y}_D )_i^{\ j} + \frac{\lambda_D}{\Lambda^3}
(E_D)_{i \alpha} (\bar{\Phi}_0^T)^{\alpha k} \left( 
( \hat{E}_0^{\prime})_k^{\ l} 
+ a_D  ( \hat{X}_2 )_k^{\ l}  \right) ({\Phi}_0)_{l \beta} (\bar{E}_D)^{\beta j}
 \right\} (\hat{\Theta}_D)_j^{\ ij} ,
\eqno(2.19)
$$
$$
W_R = \left\{  \mu_R (Y_R)_{ij}  +
{\lambda_R} \left[ (\hat{Y}_e)_i^{\ k} ({\Phi}_u)_{k j}
+ (\Phi_u)_{ik} (\hat{Y}_e^T)^k_{\ j} \right]  \right\} 
(\bar{E})^{jl} (\hat{\Theta}_R)_l^{\ i} .
\eqno(2.20)
$$
In Eq.(2.20), we have used $(\bar{E})^{jl} (\hat{\Theta}_R)_l^{\ i}$
without using $(\bar{\Theta}_R)^{ji}$. 
Although this is somewhat factitious, this was required 
in order to make flavon sector anomaly free as seen in Table 1.

On the other hand, 
 for the VEV relation as to ${\Phi}_u$ (and also
${\Phi}_d$), we assume the following superpotential 
without $\Theta$ fields:
$$
W_q = \frac{\lambda_q}{\Lambda'} {\rm Tr} \left\{ \left[ {\Phi}_u +
 \frac{\lambda_u}{\Lambda^4} 
P_u \bar{\Phi}_0 ({E}_0+ a_u {X}_3) \bar{\Phi}_0^T P_u \right]  \bar{E}
\left[ {\Phi}_d + \frac{\lambda_d}{\Lambda^4} E_d 
\bar{\Phi}_0 ({E}_0+ a_d {X}_3) \bar{\Phi}_0^T  E_d \right] \bar{E} \right\} 
$$
$$
+ \frac{\lambda_q}{\Lambda'} {\rm Tr} \left\{ \left[ \bar{\Phi}_u + 
\frac{\lambda_u}{\Lambda^4}
\bar{P}_u {\Phi}_0 (\bar{E}_0+ a_u \bar{X}_3) {\Phi}_0^T \bar{P}_u \right]  E 
\left[ \bar{\Phi}_d + \frac{\lambda_d}{\Lambda^4} \bar{E}_d 
{\Phi}_0 (\bar{E}_0+ a_d \bar{X}_3) {\Phi}_0^T  \bar{E}_d \right] E \right\} .
\eqno(2.21)
$$
SUSY vacuum conditions $\partial W/\partial {\Phi}_u =0$ and so on 
lead to the VEV relations (2.4) and (2.5).
Other conditions $\partial W/\partial \bar{\Phi}_0 =0$, 
$\partial W/\partial E_0 =0$, and so on are satisfied identically 
under the VEV relations (2.4) and (2.5). 
A reason that we have introduced the new superpotential form
(2.21) is that we need a direct $R$ charge relation between 
${\Phi}_u$ and ${\Phi}_d$ from a phenomenological reason. 
(We will discuss in Sec.2.4.)  
From the superpotential (2.21), we have the following $R$ charges relations
$$
\begin{array}{l}
R({\Phi}_u) +R({\Phi}_d) + 2 R(\bar{E})  = 2, \\
R(\bar{\Phi}_u) +R(\bar{\Phi}_d) + 2 R({E}) = 2 .
\end{array}
\eqno(2.22)
$$

\begin{table}
\caption{Assignments of 
SU(2)$_L \times$SU(3)$_c \times$U(3)$\times$U(3)$'$. 
$R$ charges are discussed in Sec.2.4.}
\begin{center}
\begin{tabular}{|c|ccccc|ccc|cc|} \hline
& $\ell$ & $e^c$ & $\nu^c$ & $N$ & $N^c$ & $q$ & $u^c$ & $u^c$ & 
$H_u$ & $H_d$  \\ \hline
SU(2)$_L$ & ${\bf 2}$ & ${\bf 1}$ & ${\bf 1}$ & ${\bf 1}$ & ${\bf 1}$ &
${\bf 2}$ & ${\bf 1}$ & ${\bf 1}$ & ${\bf 1}$ & ${\bf 1}$ \\ 
SU(3)$_c$ & ${\bf 1}$ & ${\bf 1}$ & ${\bf 1}$ & ${\bf 1}$ & ${\bf 1}$ & 
${\bf 3}$ & ${\bf 3}^*$ & ${\bf 3}^*$ & ${\bf 1}$ & ${\bf 1}$ \\ \hline
U(3) & ${\bf 3}$ & ${\bf 3}^*$ & ${\bf 3}^*$ & ${\bf 3}$ & ${\bf 3}^*$ & 
${\bf 3}$ & ${\bf 3}^*$ & ${\bf 3}^*$ & ${\bf 1}$ & ${\bf 1}$ \\
U(3)$'$ &  ${\bf 1}$ &   ${\bf 1}$ &  ${\bf 1}$ &  ${\bf 1}$ & ${\bf 1}$ & 
${\bf 1}$ &  ${\bf 1}$ &   ${\bf 1}$ &  ${\bf 1}$ &  ${\bf 1}$ 
\\  \hline
\end{tabular}

\vspace{2mm}

\begin{tabular}{|ccc|cccccc|} \hline
 $\hat{Y}_e$ & $\hat{Y}_D$ & $Y_R$ & $\hat{Y}_u$ & $\hat{Y}_d$ & 
${\Phi}_u$ &  $\bar{\Phi}_u$ & ${\Phi}_d$ &  $\bar{\Phi}_d$ 
\\ \hline
 ${\bf 1}$ & ${\bf 1}$ & ${\bf 1}$ &  
${\bf 1}$ &  ${\bf 1}$ & 
${\bf 1}$ & ${\bf 1}$ & ${\bf 1}$ & ${\bf 1}$ 
\\ 
 ${\bf 1}$ & ${\bf 1}$ & ${\bf 1}$ & 
 ${\bf 1}$ &  ${\bf 1}$ & 
${\bf 1}$ & ${\bf 1}$ & ${\bf 1}$ & ${\bf 1}$ 
\\ \hline
 ${\bf 8}+{\bf 1}$ & ${\bf 8}+{\bf 1}$ & ${\bf 6}$ & 
 ${\bf 8}+{\bf 1}$ &  ${\bf 8}+{\bf 1}$ & 
 ${\bf 6}$ & ${\bf 6}^*$ &  ${\bf 6}$ & ${\bf 6}^*$ 
 \\
 ${\bf 1}$ &   ${\bf 1}$ &  ${\bf 1}$ &  
${\bf 1}$ & ${\bf 1}$ & 
${\bf 1}$ &  ${\bf 1}$ &   ${\bf 1}$ &  ${\bf 1}$ 
\\ \hline
\end{tabular}

\vspace{2mm}

\begin{tabular}{|cccccc|cc|} \hline
${\Phi}_0$ & $\bar{\Phi}_0$ & ${E}_0$ & $\bar{E}_0$ & ${X}_3$  &
$\bar{X}_3$ & $\hat{E}'_0$ & $\hat{X}_2$  \\ \hline
${\bf 1}$ &
${\bf 1}$ & ${\bf 1}$ & ${\bf 1}$ & ${\bf 1}$ & ${\bf 1}$ &
 ${\bf 1}$ &   ${\bf 1}$ \\
 ${\bf 1}$ & ${\bf 1}$ & ${\bf 1}$ & ${\bf 1}$ & ${\bf 1}$ &  ${\bf 1}$ &
 ${\bf 1}$ & ${\bf 1}$ \\  \hline
${\bf 3}$ & ${\bf 3}^*$ & ${\bf 1}$ & ${\bf 1}$ &
${\bf 1}$ & ${\bf 1}$ & 
${\bf 8}+{\bf 1}$ & ${\bf 8}+{\bf 1}$ \\
${\bf 3}$ & ${\bf 3}^*$ & 
${\bf 6}$ &  ${\bf 6}^*$ & 
${\bf 6}$ & ${\bf 6}^*$ & ${\bf 1}$ &   ${\bf 1}$ \\ \hline
\end{tabular}

\vspace{2mm}

\begin{tabular}{|ccccccc|cc|ccccc|} \hline
 ${E}$ & $\bar{E}$ &  ${E}_D$ & $\bar{E}_D$ & ${E}_d$ & $\bar{E}_d$ & 
  $\bar{E}_N$ & 
 ${P}_u$ &  $\bar{P}_u$ &
 $\hat{\Theta}_e$ &  $\hat{\Theta}_D$ & $\hat{\Theta}_R$ & 
$\hat{\Theta}_u$ &  $\hat{\Theta}_d$   
\\ \hline
 ${\bf 1}$ &   ${\bf 1}$ &  ${\bf 1}$ & ${\bf 1}$ & ${\bf 1}$ &
 ${\bf 1}$ &   ${\bf 1}$ &  ${\bf 1}$ &  ${\bf 1}$ & 
 ${\bf 1}$ &  ${\bf 1}$ & ${\bf 1}$ & ${\bf 1}$ & ${\bf 1}$ \\
${\bf 1}$ &  ${\bf 1}$ &  ${\bf 1}$ &
 ${\bf 1}$ &   ${\bf 1}$ &  ${\bf 1}$ &  ${\bf 1}$ & 
${\bf 1}$ &  ${\bf 1}$ &  ${\bf 1}$ &  ${\bf 1}$ &  ${\bf 1}$ &
${\bf 1}$ & ${\bf 1}$ 
\\ \hline
 ${\bf 6}$ &  ${\bf 6}^*$ &  ${\bf 6}$ &  ${\bf 6}^*$ &  ${\bf 6}$ & 
 ${\bf 6}^*$ &  ${\bf 6}^*$ & 
   ${\bf 6}$ & ${\bf 6}^*$ & 
${\bf 8}+{\bf 1}$ &  ${\bf 8}+{\bf 1}$ & 
${\bf 8}+{\bf 1}$ & ${\bf 8}+{\bf 1}$ & ${\bf 8}+{\bf 1}$ \\
 ${\bf 1}$ &  ${\bf 1}$ &  ${\bf 1}$ &   ${\bf 1}$ &  ${\bf 1}$ & ${\bf 1}$ &
 ${\bf 1}$ &   ${\bf 1}$ &  ${\bf 1}$ &  ${\bf 1}$ &  ${\bf 1}$ & 
 ${\bf 1}$ &  ${\bf 1}$ & ${\bf 1}$ 
\\ \hline
\end{tabular}

\end{center}

\end{table}

In Table 1, we list all flavons in this model.
As seen in Table 1, 
the yukawaons in the lepton  and quark sectors belong to ${\bf 8}+{\bf 1}$ 
of U(3), so that sum of the anomaly coefficients 
obviously take zero both in the lepton  and quark sectors: 
$$
\begin{array}{l}
\sum_{\rm leptons} A = 3\, A({\bf 3}) + 3\, A({\bf 3}^*) = 0,  \\
\sum_{\rm quarks} A = 6\, A({\bf 3}) + 6\, A({\bf 3}^*) = 0 .
\end{array}
\eqno(2.23)
$$
Besides, sum of the anomaly coefficients becomes  zero in the flavon sector, too:
$$
\sum_{\rm flavons} A = 8\, A({\bf 6}) + 8\, A({\bf 6}^*) + 
1\, A({\bf 3}) + 1\, A({\bf 3}^*) +10 \, A({\bf 8}+{\bf 1}) = 0 .
\eqno(2.24)
$$
as seen in Table 1.
In Table 1, we can obviously see that U(3)$'$ is also anomaly free.

\vspace{2mm}

\noindent{\bf 2.3 \ Flavons with specific VEV forms}

In this subsection, we discuss flavon VEV matrices with 
specific forms, i.e. $\langle E\rangle$'s with a unit matrix, 
$\langle P_u\rangle$ with a phase matrix form, and
$\langle {X}_3 \rangle$ with a democratic form 
defined in Eq.(1.4). 
However, we do not discuss the origin of the form 
$\langle \hat{X}_2\rangle$. 
In this paper, the form has been required only 
based on a phenomenological reason, and it is purely ad hoc one.

First, we discuss $R$ charges for flavons $E$'s, 
whose VEV matrices have the same matrix forms ${\bf 1}$. 
For example, we assign $R$ charges for $E$ and $P_u$ 
as follows:
$$
\begin{array}{l}
 R(P_u) +R(\bar{E}) = 1 , \\ 
R(E) +R(\bar{P}_u) =1.
\end{array}
\eqno(2.25)
$$
(Although another choice  $R(P_u) +R(\bar{P}_u) = 
R(E) +R(\bar{E}) =1$ is possible, the choice (2.25) 
is useful for parameter from the phenomenological
point of view as seen in Eq.(2.33) later.) 
Then, we assume the following superpotential
$$
W=\lambda_1 {\rm Tr}[E \bar{E} P_u \bar{P}_u] +
\lambda_2 {\rm Tr}[E \bar{E}] {\rm Tr}[P_u \bar{P}_u] ,
\eqno(2.26)
$$
where we have neglected possible terms 
${\rm Tr}[E \bar{P}_u P_u \bar{E}]$
and ${\rm Tr}[E \bar{P}_u] {\rm Tr}[ P_u \bar{E}]$. 
This is an assumption based on the phenomenological 
requirement, so that this is only ad hoc one.
The SUSY vacuum conditions for (2.26) lead to 
$$
  \langle \bar{E} \rangle \langle {E} \rangle \propto {\bf 1} , 
\eqno(2.27)
$$
$$  
 \langle \bar{P}_u \rangle \langle {P}_u \rangle \propto {\bf 1}  .
\eqno(2.28)
$$  
Therefore, the relation (2.27) leads to 
$$
  \langle \bar{E} \rangle = \langle {E} \rangle
  = {\bf 1} ,
\eqno(2.29)
$$
for the choice (2.8), 
while the relation (2.28) leads to a phase matrix form Eq.(2.9)
when we assume that $\langle {P}_d \rangle$ is diagonal and  
a $D$ term condition with the type (2.7).

For $E_D$, $E_d$, and so on, we take $R$ charge assignments
$$
R(E_D) + R(\bar{E}_D)= 1, \ \ R(E_d) + R(\bar{E}_d)= 1, 
\eqno(2.30)
$$
and so on. 
We assume the $D$ term condition of the type (2.7) for 
superpotential forms similar to (2.26). 
Then, we can obtain the VEV matrix forms 
$\langle E_D \rangle = \langle E_d \rangle = {\bf 1}$.  

For the VEV form of the flavon $X_3$, we assume the 
following scenario:
The U(3)$'$ is broken into a permutation symmetry S$_3$ 
at an energy scale $\Lambda'$ which is larger than 
the U(3) symmetry breaking scale $\Lambda$. 
Therefore, in a transition of a field $\hat{X}_3$ into 
a VEV matrix $\langle\hat{X}_3\rangle$, i.e. 
$$
(\bar{\Phi}_0)^{i\alpha} \left[(\hat{E}_0)_\alpha^{\ \beta}
 + a_f (\Hat{X}_3)_\alpha^{\ \beta} \right]
  (\Phi_0^T)_{\beta j} \ \rightarrow \ 
(\bar{\Phi}_0)^{i\alpha} \left[ \langle \hat{E}_0 \rangle_\alpha^{\ \beta} 
+ a_f \langle \Hat{X}_3 \rangle_\alpha^{\ \beta} \right] 
  (\Phi_0^T)_{\beta j} ,
\eqno(2.31)
$$  
the factor $(\langle \hat{E}_0\rangle + a_f \langle \hat{X}_3\rangle)$ takes 
a form [(a unit matrix)
+ (a democratic matrix)],
where we have tacitly assumed that $[\hat{E}_0 + a_f \hat{X}_3]_\alpha^{\ \beta}$
takes a VEV value prior to $\langle \Phi_0 \rangle_{\alpha i}$. 

For the origin of the VEV form $\langle X_2\rangle$, 
we have no idea at present.
The form is purely ad hoc one motivated from the phenomenological point
of view.
(Some ideas on $X_2$ are found in Refs.\cite{K-N_EPJC12,K-N_EPJC13}, 
but those are still controversial.

\vspace{2mm}

\noindent{\bf 2.4 \ $R$ charge assignments}

As seen in Table 1, 
some of flavons have the same transformation properties under the 
U(3)$\times$U(3)$'$ symmetries. 
Those are distinguished by their $R$ charges. 
At preset, we cannot fix  $R$ charges of flavons uniquely, 
because we have a considerable number of flavons compared with 
the number of the VEV relations required.

When we assign $R$ charges to flavons, we must pay attention to the 
following points:
(i) Careless assignment allows unwelcome superpotential terms.  
(ii) If we allow a family singlet combination with $R=0$, 
for example ${\rm Tr}[A\bar{B}]$,  terms $\left({\rm Tr}[A\bar{B}]\right)^n$ 
with any $n$ 
can be attached to superpotential terms with $R=2$, 
so that such combination should be forbidden.

Let us demonstrate an example of $R$ charge assignments.
From Eqs.(2.4) and (2.5), we have the following constraints:
$$
\begin{array}{l}
 R({\Phi}_u) -2 R(P_u) = R({\Phi}_d) -2 R(E_d) 
= R_0 \equiv 2 R(\bar{\Phi}_0) + R({X}_3), \\
 R(\bar{\Phi}_u) -2 R(\bar{P}_u) = R(\bar{\Phi}_d) -2 R(\bar{E}_d) 
= \bar{R}_0 \equiv 2 R({\Phi}_0) + R(\bar{X}_3) .
\end{array}
\eqno(2.32)
$$
Therefore, from the constraint (2.22), we obtain
$$
\begin{array}{l}
 R({\Phi}_u) = 1 -R(\bar{E}) -R(E_d) +R(P_u) , \\
 R({\Phi}_d) = 1 -R(\bar{E}) -R(P_u) +R(E_d) , \\
\end{array}
\eqno(2.33) 
$$
which lead to
$$
\begin{array}{l}
 R({\Phi}_u) = 2 R(P_u) -R(E_d) , \\
 R({\Phi}_d) = R(E_d) , \\
\end{array}
\eqno(2.34) 
$$
respectively, 
from the $R$ charge relations (2.25).
This means that flavon 
${(\Phi}_d)_{ij}$ can mix with $(E_d)_{ij}$. 
Therefore, The VEV relations given in Eq.(2.5) must be modified as 
$$
\begin{array}{l}
\langle {\Phi}_d\rangle_{i j} = 
\langle E_d \rangle_{ik} \langle \bar{\Phi}_0 \rangle^{k\alpha} 
\left( \langle {E}_0 \rangle_{\alpha \beta} + a_d 
 \langle {X}_3 \rangle_{\alpha \beta} \right) 
\langle \bar{\Phi}_0^T\rangle^{\beta l}  \langle {E}_d \rangle_{lj} 
+ \xi_0^d \langle E_d \rangle_{ij}, \\
\langle \bar{\Phi}_d\rangle^{i j} = 
\langle \bar{E}_d \rangle^{ik} \langle {\Phi}_0 \rangle_{k\alpha} 
\left( \langle \bar{E}_0 \rangle^{\alpha \beta} + a_d 
 \langle \bar{X}_3 \rangle^{\alpha \beta} \right)  
\langle {\Phi}_0^T\rangle_{\beta l}  \langle \bar{E}_d \rangle^{lj} 
+  \xi_0^d \langle \bar{E}_d \rangle^{ij} .
\end{array}
\eqno(2.35)
$$
For Eq.(2.4) in the up-quark sector, possible additive terms such as 
(2.35) do not appear.

\begin{table}
\caption{$R$ charge assignments. 
For more relations, see Eq.(2.36).}
\begin{center}
\begin{tabular}{|ccccc|ccc|cc|} \hline
 $\ell$ & $e^c$ & $\nu^c$ & $N$ & $N^c$ & $q$ & $u^c$ & $u^c$ & 
$H_u$ & $H_d$  \\ \hline
 $r_\ell$ & $r_{ec}$ & $r_{\nu c}$ & $r_N$ & $r_{N c}$ & $r_q$ & 
$r_{u c}$ & $r_{u c}$ & $r_{Hu}$ & $r_{Hd}$ 
\\ \hline
\end{tabular}

\vspace{2mm}
\begin{tabular}{|ccc|cccccc|} \hline
 $\hat{Y}_e$ & $\hat{Y}_D$ & $Y_R$ & 
$\hat{Y}_u$ & $\hat{Y}_d$ & ${\Phi}_u$ &  $\bar{\Phi}_u$ & 
${\Phi}_d$ & $\bar{\Phi}_d$
\\ \hline
 $r_e$ & $r_D$ & $r_e +r_u$ &  $r_u +\bar{r}_u$ &  $r_d +\bar{r}_d$ & 
$r_u$ & $\bar{r}_u$ & $r_d$ & $\bar{r}_d$  
\\ \hline
\end{tabular}

\vspace{2mm}
\begin{tabular}{|cccccc|cc|} \hline
${\Phi}_0$ & $\bar{\Phi}_0$ & ${E}_0$ & $\bar{E}_0$ & 
${X}_3$  & $\bar{X}_3$ & $\hat{E}'_0$ & $\hat{X}_2$
\\ \hline
$r_0$ & $\bar{r}_0$ & $r_{X3}$ & $\bar{r}_{X3}$ & 
$r_{X3}$  & $\bar{r}_{X3}$ & $\hat{r}_{X2}$ & $\hat{r}_{X2}$
\\ \hline
\end{tabular}

\vspace{2mm}
\begin{tabular}{|ccccccc|cc|} \hline
${E}$ & $\bar{E}$ & ${E}_D$ & $\bar{E}_D$ &  ${E}_d$ & $\bar{E}_d$ & $\bar{E}_N$ & 
 ${P}_u$ &  $\bar{P}_u$   
\\ \hline
$r_{E}$ & $\bar{r}_{E}$ & $r_{ED}$ & $1 - {r}_{ED}$ & $r_{d}$ & 
$\bar{r}_{d}$ & $2-2 r_N$ & 
 $1-\bar{r}_E$ &  $1-r_E$
\\ \hline
\end{tabular}

\vspace{2mm}
\begin{tabular}{|ccccc|} \hline
 $\hat{\Theta}_e$ &  $\hat{\Theta}_D$ & $\hat{\Theta}_R$ & 
$\hat{\Theta}_u$ &  $\hat{\Theta}_d$ 
\\ \hline
 $2-r_e$ &  $2-r_D$ & $2-(\bar{r}_E +r_e +r_u)$ & 
$2-(r_u+\bar{r}_u)$ & $2-(r_d+\bar{r}_d)$  
\\ \hline
\end{tabular}
\end{center}

\end{table}

Finally,  we summarize the $R$ charge assignments in Table 2. 
These $R$ charges should satisfy the following relations:
$$
\begin{array}{l}
R(\hat{Y}_e) \equiv r_e = 2-r_\ell - r_{ec} -r_{Hd} =
(r_0 +\bar{r}_0) +(r_{X3} +\bar{r}_{X3}) , \\
R(\hat{Y}_D) \equiv r_D = 2-r_\ell - r_{\nu c} -r_{Hu} =
(r_{ED} +\bar{r}_{ED}) +(r_0 +\bar{r}_0)+ r_{X2} , \\
R(Y_R) =2-r_{\nu c} -r_{Nc} =r_e +r_u  , \\
R(\hat{Y}_u)  =2-r_q -r_{uc} -r_{Hu}= r_u + \bar{r}_u , \\
R(\hat{Y}_d)  =2-r_q -r_{dc} -r_{Hd}= r_d + \bar{r}_d , \\ 
r_u -2 r_{Pu} = r_d - 2 r_{Ed} = 2 r_0 + r_{X3},  \\
\bar{r}_u -2 \bar{r}_{Pu} = \bar{r}_d - 2 \bar{r}_{Ed} = 2 \bar{r}_0 + \bar{r}_{X3}, \\
r_u + r_d + 2 \bar{r}_E =2 , \\
\bar{r}_u + \bar{r}_d + 2 {r}_E =2 . \\
\end{array}
\eqno(2.36)
$$


\vspace{3mm}

\noindent{\large\bf 3 \ Parameter fitting}

\vspace{2mm}

\noindent{\bf 3.1 \ How many parameters?}

We summarize our mass matrices $M_f$ for the lepton sector 
($f=e$, $D$, and $\nu$) and the quark sector ($f=u$ and $d$) as follows:
$$
M_e  = \Phi_0 ( {\bf 1} + 
a_e X_3 )\Phi_0 ,
\eqno(3.1)
$$
$$
M_D = \Phi_0 ( {\bf 1} + 
a_D X_2) \Phi_0 ,
\eqno(3.2) 
$$
$$
M_u = P_u  \Phi_0  \left( {\bf 1} + 
a_u e^{i \alpha_u} X_3 \right) \Phi_0 \cdot 
\Phi_0  \left( {\bf 1} + 
a_u e^{i \alpha_u} X_3 \right) \Phi_0 P_u^\dagger ,
 \eqno(3.3)
$$
$$
M_d = \left[\Phi_0  \left( {\bf 1} + 
a_d e^{i \alpha_d} X_3 \right) \Phi_0  + \xi^d_0 {\bf 1} \right]\cdot 
\left[ \Phi_0  \left( {\bf 1} + 
a_d e^{i \alpha_d} X_3 \right) \Phi_0  + \xi^d_0 {\bf 1} \right] 
,
\eqno(3.4)
$$
$$
M_\nu = M_D Y_R^{-1} M_D \cdot M_D Y_R^{-1} M_D , \ \ \ \ 
Y_R = Y_e \Phi_u  + \Phi_u Y_e . 
\eqno(3.5)
$$
Here, for convenience, we have dropped 
the notations ``$\langle$" and ``$\rangle$". 
Since we are interested only in the mass ratios and mixings, 
we use dimensionless expressions
$\Phi_0 = {\rm diag}(x_1, x_2, x_3)$,  
$P_u= {\rm diag} (e^{-i\phi_1}, e^{-i\phi_2},1)$, 
and $E={\rm diag}(1,1,1)$. 
Therefore, the parameters $a_e$, $a_D$, $\cdots$, are re-defined 
by Eqs.(3.1)-(3.5). 
Since we have assumed that the parameters $a_f$ are real in the lepton sector, 
while those are complex in the quark sector,  
we have denoted the parameters $a_u$ and $a_d$ in the quark 
sectors as $a_u e^{i\alpha_u}$ and $a_d e^{i\alpha_d}$.

Besides, we require ``economy of the number of parameters".
We neglect parameters which play no essential roles in numerical 
fitting to the mixings and mass ratios as far as possible. 
Namely we require
$$
 \phi_1=0 ,
\eqno(3.6)
$$ 
by way of trial.

Therefore, in the present model, we have 10 adjustable parameters,  
$(x_1/x_2, x_2/x_3)$, $a_e$, $a_D$, $(a_u, \alpha_u)$, $(a_d, \alpha_d)$,
$\xi_0^d$, and $\phi_2$   
for the 18 observable quantities (8 mass ratios in the
charged lepton, up-quark-, down-quark-, and neutrino-sectors, 4 CKM 
mixing parameters, and 4+2 PMNS mixing parameters). 
In order to fix these parameters, we use, as input values, 
the observed values for ${m_e}/{m_\mu}$, ${m_\mu}/{m_\tau}$, 
${m_c}/{m_t}$, ${m_u}/{m_c}$, 
$\sin^2 2 \theta_{12}$, $R_\nu \equiv {\Delta m_{21}^2}/{\Delta m_{32}^2}$, 
${m_d}/{m_s}$,  ${m_s}/{m_b}$, $|V_{us}|$, and $|V_{cb}|$ as shown later. 
The process of fixing parameters are summarized in Table. 3. 
The parameter fitting will be done quantitatively 
(not qualitatively). 
Observed values which should be fitted are values at 
$\mu = M_Z$.  

Note that the purpose of the present paper is not 
to compete with other models for reducing parameter number 
in the model, but it is to investigate whether it is 
possible or not to fit all of the mixing parameters and 
mass ratios without using any family number dependent 
parameters when we use only the observed charged lepton 
masses as family dependent parameters.
If we pay attention only to fitting of mixing parameters, 
a model with fewer number of parameters based on quark-lepton complementarity
\cite{q-l_compl} is rather 
excellent compared with the preset model.
(For such a recent work, for example, see 
Ref.\cite{q-l_compl_PRD12} and references there in.)

\begin{table}
\caption{Process for fitting parameters. 
}
\vspace{2mm}
\begin{center}
\begin{tabular}{|c|cc|cc|c|} \hline
Step & Inputs & $N_{inp}$ &  Parameters & $N_{par}$ &
 Predictions  \\ \hline
1st  & $\frac{m_e}{m_\mu}$, $\frac{m_\mu}{m_\tau}$, 
$\frac{m_u}{m_c}$, $\frac{m_c}{m_t}$ & 5
& $\frac{x_1}{x_2}$, $\frac{x_2}{x_3}$, $a_e$ & 5 &    \\
    &  $\sin^2 2\theta_{12}$ &  & $a_u$, $\alpha_u$ & &  
\\ \hline
2nd  & $R_\nu$  & 1 & $a_D$ & 1 &
$\sin^2 2\theta_{13}$, $\sin^2 2\theta_{23}$, $\delta_{CP}^\ell$  \\
    &    &   &   &  & 2 Majorana phases, $\frac{m_{\nu 1}}{m_{\nu 2}}$,
 $\frac{m_{\nu 2}}{m_{\nu 3}}$   \\ \hline
3rd  & $\frac{m_s}{m_b}$, $\frac{m_d}{m_s}$, $|V_{us}|$, &
 3 & $a_d$, $\alpha_d$, $\xi_0^d$ & 3 &  \\ \hline
4th  & $|V_{cb}|$  & 1 &   $\phi_2$ & 1 & 
$|V_{ub}|$, $|V_{td}|$, $\delta_{CP}^q$  \\ \hline
option &  $\Delta m^2_{32}$ &   & $m_{\nu 3}$ &  & 
$(m_{\nu 1}, m_{\nu 2},  m_{\nu 3})$, $\langle m \rangle$  \\
\hline
$\sum N_{\dots}$ & & 10 &  & 10 &   \\ \hline 
\end{tabular}
\end{center}
\end{table}

\vspace{2mm}

\noindent{\bf 3.2 \ PMNS mixing}

Let us present the details of parameter fitting to the PMNS mixings. 
Under a given $a_e$, the relative ratios of parameters $(x_1, x_2, x_3)$ in $\Phi_0$ are fixed  
by the ratios of the charged lepton masses ${m_e}/{m_\mu}=0.004738$ and
 ${m_\mu}/{m_\tau}=0.05883$. 
Since the mass ratios of the up quarks and the lepton mixing parameter 
$\sin^2 2\theta_{12}$ depends only on $a_e$ and $(a_u, \alpha_u)$, 
we first fix the following 
parameter values of $a_e$ and $(a_u, \alpha_u)$ 
$$
(a_e, a_u, \alpha_u) \sim (8.0, -1.273, -1.4^\circ) ,
\eqno(3.7)
$$
which are fixed 
from the observed values of $m_c/m_t$, $m_u/m_c$, and
$\sin^2 2\theta_{12}$:
$$
r^u_{12} \equiv \sqrt{\frac{m_u}{m_c}} 
= 0.045^{+0.013}_{-0.010} , \ \ \ \ 
r^u_{23} \equiv \sqrt{\frac{m_c}{m_t}}
=0.060 \pm 0.005 ,
\eqno(3.8)
$$
at $\mu=m_Z$ \cite{q-mass}, and
$\sin^2 2\theta_{12} = 0.857 \pm 0.024$ \cite{PDG12}. 
(These values will be fine-tuned in whole parameter
fitting of $U_{PMNS}$ and $V_{CKM}$ later.)
The parameters $(x_1/x_2, x_2/x_3)$  are fixed as ($0.07300$, $0.3825$). 
Note that we do not change the mass matrix structures 
for $M_e$, $M_u$, and $M_\nu$ from the previous paper
\cite{K-N_PLB12}. However $M_d$ is different 
from the previous model, so that we refit these parameters 
in order to reproduce the observed CKM mixing parameters too 
as seen later.

\begin{figure}[h]
\begin{picture}(200,200)(0,0)

  \includegraphics[height=.3\textheight]{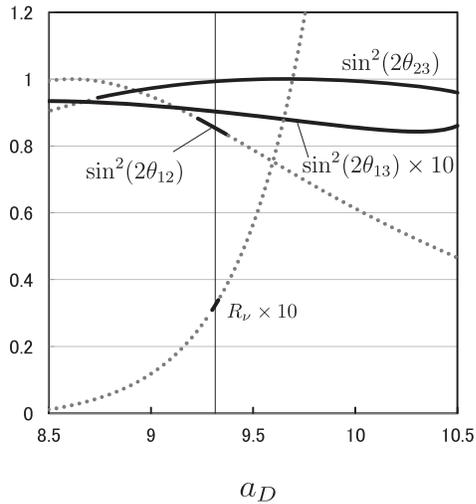}

\end{picture}  
  \caption{Lepton mixing parameters $\sin^2 2\theta_{12}$, 
$\sin^2 2\theta_{23}$, $\sin^2 2\theta_{13}$, and the neutrino 
mass squared difference ratio $R_\nu$ versus the parameter $a_D$. 
We draw curves of the lepton mixing parameters as functions of $a_D$, with taking  
$a_e=8.0$, $a_u=-1.273$, and $\alpha_u=-1.4^\circ$.
The solid and dotted parts of the curves are, respectively, within and 
out of the observed ranges given by (3.10)-(3.13). We find that the parameter 
$a_D$ around $a_D=9.32$ is consistent with all the observed values 
\cite{PDG12}.  
}
  \label{a_D}
\end{figure}

In the present model, lepton mixing parameters depend only the parameter $a_D$ 
after we fix $a_e$ and $(a_u, \alpha_u)$ as (3.7). 
We illustrate the behaviors of lepton mixing parameters 
$\sin^2 2\theta_{12}$, $\sin^2 2\theta_{23}$, 
$\sin^2 2\theta_{13}$, and the neutrino 
mass squared difference ratio 
${R_{\nu}}$ versus the parameter $a_D$. 
We draw curves of the lepton mixing parameters and $R_\nu$ 
as functions of $a_D$ in Fig. 1.
As seen in Fig.1, the predictions of $\sin^2 2\theta_{12}$ and 
$R_{\nu}$ are 
sensitive to the parameter $a_D$, while  
the prediction of $\sin^2 2 \theta_{23}$ and $\sin^2 2 \theta_{13}$
are insensitive to $a_D$.
Using Fig. 1, we do fine tuning of the parameter $a_D$ as
$$
a_D=9.32, 
\eqno(3.9)
$$
in order to fit the observed values \cite{PDG12} given by
$$
\sin^2 2\theta_{12} = 0.857 \pm 0.024,
\eqno(3.10)
$$
$$
R_{\nu} \equiv \frac{\Delta m_{21}^2}{\Delta m_{32}^2}
=\frac{m_{\nu2}^2 -m_{\nu1}^2}{m_{\nu3}^2 -m_{\nu2}^2}
=\frac{(7.50\pm 0.20) \times 10^{-5}\ {\rm eV}^2}{
(2.32^{+0.12}_{-0.08}) \times 10^{-3}\ {\rm eV}^2} = 
(3.23^{+0.14}_{-0.19} ) \times 10^{-2} ,
\eqno(3.11)
$$
$$
\sin^2 2\theta_{23} >0.95,
\eqno(3.12)
$$
and
$$
\sin^2 2\theta_{13} =0.095\pm0.010. 
\eqno(3.13)
$$

\vspace{2mm}

\noindent{\bf 3.3 \ CKM mixing}

Next, we discuss quark sector. 
Since we have fixed the four parameters $a_e$, $a_u$, $\alpha_u$, and 
$a_D$, we have remaining four parameters $a_d$, $\alpha_d$, $\xi^d_0$, and 
$\phi_2$ for
eight observables (2 down-quark mass ratios and 4+2 CKM mixing 
parameters). 
The following parameters $a_d$, $\alpha_d$, and $\xi^d_0$ 
$$
a_d=-1.338, \quad \alpha_d=-14.3^\circ, \quad  \xi^d_0=0.0147
\eqno(3.14)
$$
are fixed  to fit  
the observed down-quark mass ratios at $\mu=m_Z$ \cite{q-mass}
$$
r^d_{23} \equiv \frac{m_s}{m_b} = 0.019^{+0.006}_{-0.006} , \ \ \ 
r^d_{12} \equiv \frac{m_d}{m_s} = 0.053^{+0.005}_{-0.003} ,
\eqno(3.15)
$$
and the observed CKM  mixing matrix element \cite{PDG12} 
$$
|V_{us}|=0.2252 \pm 0.0009.
\eqno(3.16)
$$
Therefore, all the CKM mixing parameters are described  
only by one remaining parameter $\phi_2$. 
We draw curves of the CKM mixing matrix elements as functions of $\phi_2$ in Fig. 2.

\begin{figure}[h]
\begin{picture}(200,200)(0,0)

  \includegraphics[height=.33\textheight]{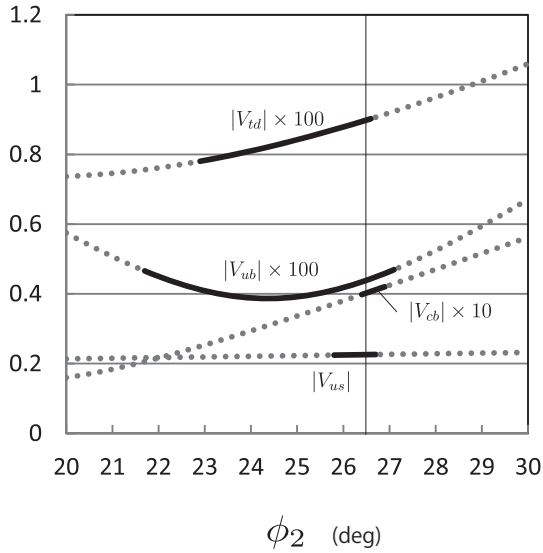}

\end{picture}  
  \caption{CKM mixing matrix elements  $|V_{us}|$, 
$|V_{cb}|$, $|V_{ub}|$, and  $|V_{td}|$ versus the parameter $\phi_2$. 
We draw curves of the CKM mixing matrix elements as functions of $\phi_2$, 
with taking  $a_e=8.0$, $a_u=-1.273$, $\alpha_u=-1.4^\circ$, $a_d=-1.338$, 
 $\alpha_d=-14.3^\circ$, and $\xi_d=0.0147$. 
The solid and dotted parts of the curves are, respectively, within and 
out of the observed ranges given by (3.16) and (3.18). 
We find that the parameter $\phi_2$ around $\phi_2=26.5^\circ$ is 
consistent with all the observed values \cite{PDG12}.  
}
  \label{phi_2}
\end{figure}

As shown in Fig.~2, all the experimental constraints on 
CKMs are satisfied by 
fine tuning the parameter $\phi_2$ around 
$$
\phi_2=26.5^\circ. 
\eqno(3.17)
$$
Here we use the values for the other observed CKM  mixing matrix 
elements \cite{PDG12} given by 
$$
|V_{cb}|=0.0409 \pm 0.0011, \ \ \
|V_{ub}|=0.00415 \pm 0.00049, \ \ \
|V_{td}|=0.0084 \pm 0.0006.
\eqno(3.18)
$$

\vspace{2mm}

\noindent{\bf 3.5 \ Summary of the parameter fitting}

Finally, we do fine-tuning of whole parameter values 
in order to give more improved fitting with the whole data. 
Our final result is as follows: 
under the parameter values
$$
a_e=8.0, \ (a_u, \alpha_u) =(-1.273,-1.4^\circ), 
\ (a_d, \alpha_d) =(-1.338, -14.3^\circ), \ \xi^d_0=0.0147,
$$
$$
\ a_D=9.32, \ \phi_2=26.3^\circ,   
\eqno(3.19)
$$
we obtain 
$$
r^u_{12} =0.0358, \ \ \ r^u_{23}= 0.0599 , \ \ \ 
r^d_{12} = 0.0547, \ \ \ r^d_{23}=0.0129,
\eqno(3.20)
$$
$$
\sin^2 2\theta_{23}=0.993, \ \ \ 
\sin^2 2\theta_{12}= 0.852, \ \ \
\sin^2 2\theta_{13} = 0.0903 , \ \ \ 
R_\nu = 0.0329,
\eqno(3.21)
$$
$$
\delta_{CP}^{\ell}= 179^\circ  \ \ \ (J^{\ell} = 6.3 \times 10^{-4} ), 
\eqno(3.22)
$$
$$
|V_{us}|=0.2256, \ \ \ |V_{cb}|=0.0402, \ \ \ 
|V_{ub}|=0.00439, \ \ \ |V_{td}|= 0.00898 ,
\eqno(3.23)
$$
$$
\delta_{CP}^{q}= 75.1^\circ   \ \ \ (J^{q} = 3.7 \times 10^{-5} ). 
\eqno(3.24)
$$
Here, $\delta^\ell_{CP}$ and $\delta^q_{CP}$ are Dirac $CP$ 
violating phases in the standard conventions of $U_{PMNS}$ 
and $V_{CKM}$, respectively. 

It should be noted that our prediction $\sin^2 2\theta_{13} = 0.0903$ is 
well consistent with the observed value in (3.13). 
Also our prediction $\sin^2 2\theta_{23}=0.993$ is roughly 
consistent with recently observed values 
$\sin^2 2\theta_{23} =0.950^{+0.035}_{-0.036}$ and 
$\sin^2 2\bar{\theta}_{23} =0.97^{+0.03}_{-0.08}$ by MINOS \cite{MINOS13}. 
Our model predicts $\delta_{CP}^{\ell}= 179^\circ$ which indicates small $CP$ 
violating effect in the lepton sector.  
Note that a recent global analysis  \cite{Fogli} has suggested that the best fit 
value for $\delta_{CP}^{\ell}$ is $1.1 \pi$.

We can also predict neutrino masses, for the parameters given by (3.19),
$$
m_{\nu 1} \simeq 0.0011\ {\rm eV}, \ \ m_{\nu 2} \simeq 0.0090 \ {\rm eV}, 
\ \ m_{\nu 3} \simeq 0.0499 \ {\rm eV}  ,
\eqno(3.25)
$$
by using the input value \cite{MINOS13}
$\Delta m^2_{32}\simeq 0.00241$ eV$^2$.
We also predict the effective Majorana neutrino mass \cite{Doi1981} 
$\langle m \rangle$ 
in the neutrinoless double beta decay as
$$
\langle m \rangle =\left|m_{\nu 1} (U_{e1})^2 +m_{\nu 2} 
(U_{e2})^2 +m_{\nu 3} (U_{e3})^2\right| 
\simeq 4.6 \times 10^{-3}\ {\rm eV}.
\eqno(3.26)
$$

We show our numerical results (predictions vs. 
observed values) in Table 4.

\begin{table}
\caption{Predicted values vs. observed values. 
}

\vspace*{2mm}
\hspace*{-6mm}
\begin{tabular}{|c|ccccccccc|} \hline
  & $|V_{us}|$ & $|V_{cb}|$ & $|V_{ub}|$ & $|V_{td}|$ & 
$\delta^q_{CP}$ &  $r^u_{12}$ & $r^u_{23}$ & $r^d_{12}$ & $r^d_{23}$ 
 \\ \hline 
Pred &$0.2256$ & $0.0402$ & $0.00439$ & $0.00898$ & $75.1^\circ$ & 
$0.0358$ & $0.0599$ & $0.0547$ & $0.0129$ 
 \\
Obs & $0.2252$ & $0.0409$ &  $0.00415$  & $0.0084$  & $68^\circ$ &
$0.045$ & $0.060$ & $0.053$  & $0.019$  
  \\ 
    &  $ \pm 0.0011$ &  $ \pm 0.0009$ & $ \pm 0.0006$ &  $ \pm 0.00049$ &
 $^{+10^\circ}_{-11^\circ}$ &
${}^{+0.013}_{-0.010}$ & $ \pm 0.005$ & $^{+0.005}_{-0.003}$ &
${}^{+0.006}_{-0.006}$ 
 \\ \hline
   & $\sin^2 2\theta_{12}$ & $\sin^2 2\theta_{23}$ & $\sin^2 2\theta_{13}$ & 
 $R_{\nu}\ [10^{-2}]$ &  
$\delta^\ell_{CP}$ & $m_{\nu 1}\ [{\rm eV}]$ & $m_{\nu 2}\ [{\rm eV}]$ & 
$m_{\nu 3}\ [{\rm eV}]$ & $\langle m \rangle \ [{\rm eV}]$ \\ \hline
 Pred & $0.852$ & $0.993$ & $0.0903$ &  $3.29$ & $179^\circ$ &
 $0.0011$ & $0.0090$ & $0.0499$ & $0.0046$ \\
Obs & $0.857$   & $ >0.95$ & $0.095$ &   $3.23 $    & -
  &  -  &  -  &  -  &  $<\mathrm{O}(10^{-1})$   \\ 
    & $ \pm 0.024$ &   & $\pm0.010$ & ${}^{+0.14}_{-0.19} $  &  &
   &    &    &    \\ \hline 
\end{tabular}
\end{table}

\vspace{3mm}

\noindent{\large\bf 4 \ Concluding remarks}

As we emphasis at Sec.1.2 and the end of Sec.3.1, 
the purpose of the present paper is not 
to compete with other models for reducing the number of parameters 
in the model.
The purpose  is to investigate whether it is 
possible or not to fit all of the mixing parameters and 
mass ratios without using any family number dependent 
parameters when we use only the observed charged lepton 
masses as family dependent parameters.
Regrettably, the answer to the above query is negative 
at present.
In order to fit all data concerned with mixings and mass 
ratios of quarks and leptons completely, we have needed 
a family-dependent matrix form $X_2$ defined Eq.(1.9) 
(with a parameter $a_D$) and a family-dependent 
parameter $\phi_2$ in the phase matrix $P_u$.
It is an open question at present whether these family-dependent 
parameters (i.e. $a_D$ and $\phi_2$) except for
charged lepton masses are indispensable or not. 
Further investigation based on a different idea will 
be required.  

We have made a revision of the yukawaon model as follows: 
(i) We have assigned the yukawaons $Y_f$ to ${\bf 8}+{\bf 1}$ of
U(3) family symmetry (not  ${\bf 6}^*$ as adopted in the
previous models), so that 
the model can become anomaly free in U(3), i.e. 
$\sum A({\rm lepton})=0$, $\sum A({\rm quark})=0$ and 
$\sum A({\rm flavon})=0$. 
(ii) Mass matrices, not only $M_u$ but also $M_d$, are 
given with a bilinear form $[\Phi_0 ({\bf 1} + a_f X_3) \Phi_0]^2$,
so that the parameter fitting has been renewed thoroughly. 
(iii) The neutrino mass matrix $M_\nu$ has been given by
a triplicate seesaw (inverse seesaw). 
(iv) In this paper, we did not refer to the origin of 
the VEV matrix form $X_2$, (1.9).
We will leave this problem to a future task.

As a result of new parameter fitting, we have obtained
the following phenomenological results:
(i) We can still obtain reasonable mass ratios and quark and 
lepton mixings, in spite of reducing the number of free parameters 
compared with the previous yukawaon model. 
(ii) For the $CP$ violation parameter in the lepton sector, 
$\delta^\ell_{CP}$, we have predicted $\delta^\ell_{CP} \simeq \pi$,
so that the $CP$ violation in the lepton sector is very small.
(iii) We have obtained an almost maximal mixing 
$\sin^2 2\theta_{23} = 0.99$ in spite of obtaining a sizable value 
of $\sin^2 2\theta_{13} = 0.09$. 
Our predicted value exists on the upper value with one $\sigma$ of
the recent observed value by MINOS \cite{MINOS13}, 
$\sin^2 2\theta_{23} =0.950^{+0.035}_{-0.036}$.
We expect that the data will be refined in the near future.  

Phenomenological success in the present work 
may support our ambitious idea that the observed hierarchical 
structure in the family mixings and mass ratios of 
quarks and leptons are caused by only one common origin,
i.e. by accepting the observed 
charged lepton mass ratios.  
However, we have still left many open questions. 
We will need further investigation in order to realize 
our goal.

\vspace{5mm}

{\large\bf Acknowledgment}

The authors thank T.~Yamashita for helpful comments on 
$D$ term conditions and the VEV forms.

%

\end{document}